\documentclass[aps,pra,showpacs,groupedaddress,floatfix,nofootinbib]{revtex4}

\usepackage{amsmath}
\usepackage{graphicx}
\usepackage{color}
\usepackage{dcolumn}

\begin{document}

\title{Comment on:``Aharonov-Bohm effect for bound states from the interaction of
the magnetic quadrupole moment of a neutral particle with axial fields''.
Phys. Rev. A \textbf{101}, 032102 (2020)}
\author{Francisco M. Fern\'andez}

\affiliation{INIFTA, Blvd. 113 y 64 S/N, Sucursal 4, Casilla de
Correo 16, 1900 La Plata, Argentina}

\begin{abstract}
We analyze recently published results about the Aharonov-Bohm
effect for bound states from the interaction of the magnetic
quadrupole moment of a neutral particle with axial fields. We show
that the eigenvalues obtained by the authors from an arbitrary
truncation of the Frobenius power series correspond to more than
one model and, consequently, are unsuitable for drawing any sound
physical conclusion. Besides, the prediction of allowed oscillator
frequencies is a mere consequence of the truncation condition just
mentioned.
\end{abstract}


\maketitle

In a recent paper Vieira and Bakke\cite{VB20} discussed the
Aharanov-Bohm effect for the bound states of a neutral particle
with a magnetic quadrupole moment that interacts with axial
fields. The Schr\"{o}dinger equation is separable in cylindrical
coordinates and the authors applied the Frobenius method to the
eigenvalue equation for the radial part. Since the expansion
coefficients satisfy a three-term recurrence relation one can, in
principle, obtain exact eigenvalues and eigenfunctions from a
straightforward truncation condition. They showed that the
eigenvalues depend on the geometric quantum phase, which gives
rise to an analog of the Ahranov-Bohm effect. In addition, they
concluded that each radial mode yields a different set of allowed
values of the angular frequency of the harmonic term included in
the interaction of one of the models. In this Comment we analyze
to which extent the truncation condition affects the physical
conclusions drawn in that paper.

The starting point of present discussion is the eigenvalue equation for the
radial part of the Schr\"{o}dinger equation
\begin{eqnarray}
&&f^{\prime \prime }(y)+\frac{1}{y}f^{\prime }(y)-\frac{\gamma ^{2}}{y^{2}}%
f(y)+\frac{\alpha }{y}f(y)-y^{2}f(y)+Wf(y)=0,  \nonumber \\
&&\gamma ^{2}=\left( l+\frac{\phi _{1}}{2\pi }\right) ^{2},\;W=\frac{\zeta
^{2}}{m\omega },\;\zeta ^{2}=2m\mathcal{E}-k^{2},\;\alpha =\frac{MB_{0}}{%
\sqrt{m\omega }},  \label{eq:eigen_eq}
\end{eqnarray}
where $l=0\pm 1,\pm 2,\ldots $ is the angular momentum quantum
number, $m$ the mass of the particle, $\omega $ the oscillator
frequency, $\mathcal{E}$ the energy, $\phi _{1}$ the geometric
quantum phase, $B_{0}$ a constant in the current density and $M$
the magnitude of the non-null components of the magnetic
quadrupole moment. The constant $-\infty <k<\infty $ comes from
the fact that the motion is unbounded along the $z$ axis;
therefore the spectrum is continuous an bounded from below
$\mathcal{E}\geq \frac{\zeta ^{2}}{2m}$. The authors simply set
$\hbar =1$, $c=1$ though there are well known procedures for
obtaining suitable dimensionless equations in a clearer and more
rigorous way\cite{F20}. In any case, $y$, $\gamma$, $\alpha$ and
$W$ are dimensionless quantities. In what follows we focus on the
discrete values of $W$ corresponding to the bound-state solutions
of equation (\ref {eq:eigen_eq}) that satisfy
\begin{equation}
\int_{0}^{\infty }\left| f(y)\right| ^{2}y\,dy<\infty .
\label{eq:bound_states}
\end{equation}
Since the behaviour of $f(y)$ at origin is determined by the term $\gamma
^{2}/y^{2}$ and its behaviour at infinity by the harmonic term $y^{2}$ we
conclude that the eigenvalue equation (\ref{eq:eigen_eq}) supports bound
states for all $-\infty <\alpha <\infty $. Besides, the eigenvalues $W$
satisfy
\begin{equation}
\frac{\partial W}{\partial \alpha }=-\left\langle \frac{1}{y}\right\rangle
<0,  \label{eq:HFT}
\end{equation}
according to the Hellmann-Feynman theorem\cite{G32,F39}. Since
$W(\alpha )$ is a continuous function of $\alpha $ it is clear
that the \textit{allowed values of the angular frequency} $\omega$
were fabricated by Vieira and Bakke by means of the truncation
method that we discuss in below.

In order to solve the eigenvalue equation (\ref{eq:eigen_eq}) the authors
proposed the ansatz
\begin{equation}
f(y)=y^{s}\exp \left( -\frac{y^{2}}{2}\right) H(y),\;H(y)=\sum_{j=0}^{\infty
}a_{j}y^{j},\;s=|\gamma |,  \label{eq:ansatz}
\end{equation}
and derived the three-term recurrence relation
\begin{eqnarray}
a_{j+2} &=&-\frac{\alpha }{\left( j+2\right) \left[ j+2\left( s+1\right)
\right] }a_{j+1}+\frac{\left( 2j-\nu \right) }{\left( j+2\right) \left[
j+2\left( s+1\right) \right] }a_{j},\;  \nonumber \\
\nu &=&W-2s-2,\;j=-1,0,\ldots ,\;a_{-1}=0,\;a_{0}=1.  \label{eq:TTRR}
\end{eqnarray}
If the truncation condition $a_n \neq 0$, $a_{n+1}=a_{n+2}=0$ has physically
acceptable solutions then one obtains some exact eigenvalues and
eigenfunctions. The reason is that, under such condition, $a_{j}=0$ for all $%
j>n$ and the factor $H(y)$ in equation (\ref{eq:ansatz}) reduces to a
polynomial of degree $n$. This truncation condition is equivalent to $\nu
=2n $ and $a_{n+1}=0$. The latter equation is a polynomial function of $%
\alpha $ of degree $n+1$ and it can be proved that all the roots $\alpha
_{s}^{(n,i)}$, $i=1,2,\ldots ,n+1$, are real\cite{CDW00,AF20}. For
convenience, we order them as $\alpha _{s}^{(n,i)}<\alpha _{s}^{(n,i+1)}$.
If $V(\alpha ,y)=-\alpha /y+y^{2}$ denotes the parameter-dependent potential
for the model discussed here, then it is clear that the truncation condition
produces an eigenvalue $W_{s}^{(n)}=2(n+s+1)$ that is common to $n+1$
different potential-energy functions $V_{s}^{(n,i)}(y)=V\left( \alpha
_{s}^{(n,i)},y\right) $. Notice that in this analysis we have deliberately
omitted part of the interaction that has been absorbed into $\gamma $ (or $s$%
) because its value is not affected by the truncation approach. Before
proceeding with the discussion of this approach we want to point out that
the truncation condition only yields \textit{some particular} eigenvalues
and eigenfunctions because not all the solutions $f(y)$ satisfying equation (%
\ref{eq:bound_states}) have polynomial factors $H(y)$. This point
will be made clearer in the discussion below. From now on, we will
refer to them as follows
\begin{equation}
f_{s}^{(n,i)}(y)=y^{s}H_{s}^{(n,i)}(y)\exp \left(
-\frac{y^{2}}{2}\right)
,\;H_{s}^{(n,i)}(y)=\sum_{j=0}^{n}a_{j,s}^{(n,i)}y^{j}.
\label{eq:f^(n,i)(y)}
\end{equation}
Present notation takes into account the multiple roots $\alpha _{s}^{(n,i)}$
mentioned above that Vieira and Bakke\cite{VB20} appeared to have overlooked.

The authors showed that $H(y)$ is a solution to a biconfluent Heun
equation but they did not use its properties to obtain their
results and resorted to the Frobenius method and the truncation
condition just outlined. For this reason we do not discuss this
equation here.

Let us consider the first cases as illustrative examples. When $n=0$ we have
$\nu =0$, $\alpha _{s}^{(0)}=0$ and the eigenfunction $f_{s}^{(0)}(y)$ has
no nodes. We may consider this case trivial because the problem reduces to
the exactly solvable harmonic oscillator. Probably, for this reason it was
not explicitly considered by Vieira and Bakke\cite{VB20}. However, the
results for $\alpha=0$ are also useful for understanding the distribution of
the eigenvalues given by the truncation method.

When $n=1$ there are two roots $\alpha _{s}^{(1,1)}=-\sqrt{4s+2}$ and $%
\alpha _{s}^{(1,2)}=\sqrt{4s+2}$ and the corresponding non-zero coefficients
are
\begin{equation}
a_{1,s}^{(1,1)}=\frac{\sqrt{2}}{\sqrt{2s+1}},\;a_{1,s}^{(1,2)}=-\frac{\sqrt{2%
}}{\sqrt{2s+1}},  \label{eq:a^(1,i)_1}
\end{equation}
respectively. We appreciate that the eigenfunction $f_{s}^{(1,1)}(y)$ is
nodeless and $f_{s}^{(1,2)}(y)$ has one node.

When $n=2$ the results are
\begin{eqnarray}
\alpha _{s}^{(2,1)} &=&-2\sqrt{4s+3},\;a_{1,s}^{(2,1)}=\frac{2\sqrt{4s+3}}{%
2s+1},\;a_{2,s}^{(2,1)}=\frac{2}{2s+1},  \nonumber \\
\alpha _{s}^{(2,2)} &=&0,\;a_{1,s}^{(2,2)}=0,\;a_{2,s}^{(2,2)}=-\frac{1}{s+1}%
,  \nonumber \\
\alpha _{s}^{(2,3)} &=&2\sqrt{4s+3},\;a_{1,s}^{(2,3)}=-\frac{2\sqrt{4s+3}}{%
2s+1},\;a_{2,s}^{(2,3)}=\frac{2}{2s+1}.  \label{eq:alpha,a_n=2}
\end{eqnarray}
In this case $f_{s}^{(2,1)}(y)$, $f_{s}^{(2,2)}(y)$ and $f_{s}^{(2,3)}(y)$
have zero, one and two nodes, respectively, in the interval $0< y<\infty $%
. Since Vieira and Bakke overlooked this multiplicity of solutions they
missed the actual meaning of the results produced by the truncation
approach. For example, one should not forget that these three functions are
states of three \textit{different} quantum mechanical models.

From the results for $n=1$ the authors derived the following equation for
the energy
\begin{eqnarray}
\mathcal{E}_{1,l,k} &=&\omega _{1,l}\left( s+2\right) +\frac{k^{2}}{2m},
\nonumber \\
\omega _{1,l} &=&\frac{M^{2}B_{0}^{2}}{2m(2s+1)},  \label{eq:E_VB}
\end{eqnarray}
and stated that ``Therefore, not all values of the angular frequency are
allowed for a polynomial of first degree''. It is clear that $\mathcal{E}%
_{1,l,k}$ is an eigenvalue common to the pair of models given by $%
V_{s}^{(1,i)}(y)$, $i=1,2$, while $\mathcal{E}_{1,l^{\prime },k}$
is an eigenvalue common to a different pair of models given by
$V_{s^{\prime }}^{(1,i)}(y)$. Therefore, it is not clear that the
results reported by Vieira and Bakke may be useful from a physical
point of view. These authors did not appear to understand the
actual meaning of the exact solutions to conditionally solvable
quantum-mechanical models that one obtains for particular
relationships about the model parameters (see, for example, \cite
{CDW00,AF20,F20b,F20c}, and in particular the remarkable
review\cite{T16}, and references therein).

As stated above, the eigenvalue equation (\ref{eq:eigen_eq}) is
not exactly
solvable; therefore, in order to obtain its eigenvalues $W_{j,s}$, $%
j=0,1,\ldots $, $W_{j,s}<W_{j+1,s}$ we should resort to a suitable
numerical method. The simplest one appears to be the well known
Rayleigh-Ritz variational method that is known to yield
increasingly accurate upper bounds to the eigenvalues of a
quantum-mechanical model\cite{P68,D20}. For present
application we choose the non-orthogonal basis set of Gaussian functions $%
\left\{ u_{j,s}(y)=y^{s+j}\exp \left( -\frac{y^{2}}{2}\right)
,\;j=0,1,\ldots \right\} $.

For simplicity we arbitrarily choose $s=0$ as a first illustrative case.
When $\alpha =\alpha _{0}^{(1,1)}=-\sqrt{2}$ the first four eigenvalues are $%
W_{0,0}=W_{0}^{(1)}=4$, $W_{1,0}=7.693978891$, $W_{2,0}=11.50604238$, $%
W_{3,0}=15.37592718$; on the other hand, when $\alpha =\alpha _{0}^{(1,2)}=%
\sqrt{2}$ we have $W_{0,0}=-1.459587134$, $W_{1,0}=W_{0}^{(1)}=4$, $%
W_{2,0}=8.344349427$, $W_{3,0}=12.53290130$. The rate of
convergence of the approach (in terms of the number $N$ of basis
functions) is clearly shown in tables \ref{tab:alpha1} and
\ref{tab:alpha2}. Notice that the truncation condition yields only
the ground state for the former model and the first excited state
for the latter, missing all the other eigenvalues for each model
potential. Compare these numerical results with the discussion
about equation (\ref{eq:a^(1,i)_1}).

As a second example we choose $s=1$, again to facilitate the calculations.
When $\alpha =\alpha _{1}^{(1,1)}=-\sqrt{6}$ the first four eigenvalues are $%
W_{0,0}=W_{1}^{(1)}=6$, $W_{1,1}=9.805784090$, $W_{2,1}=13.66928892$, $%
W_{3,1}=17.56601881$; on the other hand, when $\alpha =\alpha _{1}^{(1,2)}=%
\sqrt{6}$ we have $W_{0,1}=1.600357154$, $W_{1,1}=W_{1}^{(1)}=6$, $%
W_{2,1}=10.21072810$, $W_{3,1}=14.35078474$. Notice that the truncation
condition yields only the lowest state for the former model and the
second-lowest one for the latter, missing all the other eigenvalues for each
model potential as in the preceding example.

\begin{figure}[tbp]
\begin{center}
\includegraphics[width=9cm]{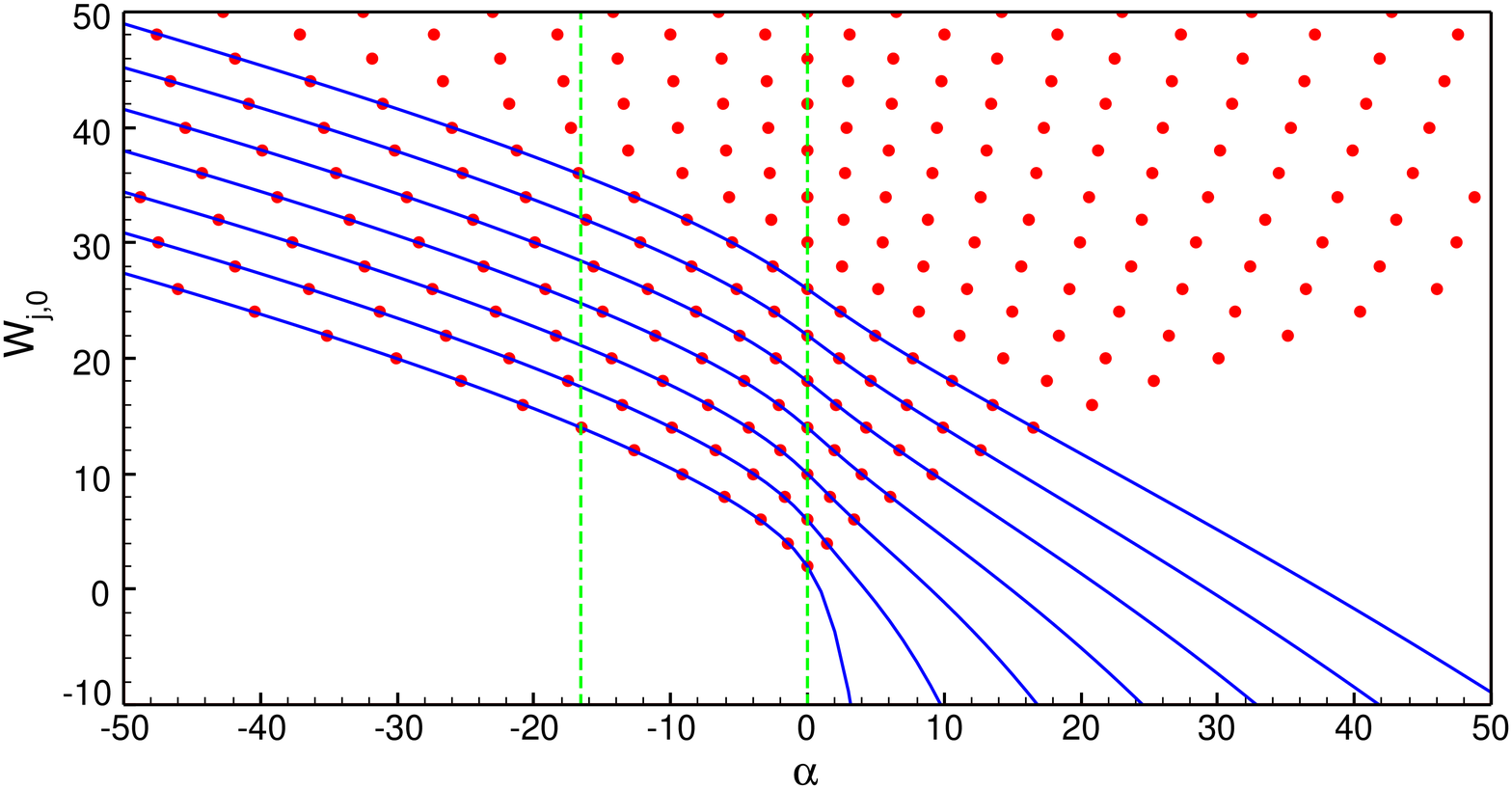}
\end{center}
\caption{Eigenvalues $W_{j,0}$ obtained from the truncation condition (red
circles) and from the variational method (blue lines)}
\label{Fig:Wn0}
\end{figure}

From the analysis above one may draw the wrong conclusion that the
truncation condition is utterly useless; however, it has been shown that one
can extract valuable information about the spectrum of conditionally
solvable models if one arranges and connects the roots $W_{s}^{(n)}$ properly%
\cite{CDW00,AF20,F20b,F20c}. From the analysis outlined above we conclude
that $\left( \alpha _{s}^{(n,i)},W_{s}^{(n)}\right) $ is a point on the
curve $W_{i-1,s}(\alpha )$, $i=1,2,\ldots ,n+1$, so that we can easily
construct some parts of such spectral curves. For example, Figure~\ref
{Fig:Wn0} shows several eigenvalues $W_{0}^{(n)}$ given by the truncation
condition (red circles) and the variational results obtained by means of $%
N=16$ Gaussian functions (blue lines). We appreciate that the
variational results connect the roots of the truncation method.
Besides, it is clear that the actual eigenvalues $W_{j,s}(\alpha
)$ are continuous functions of $\alpha $ and that the allowed
discrete angular frequencies were fabricated by Vieira and Bakke
by means of the truncation method that yields some eigenvalues for
particular values of $\alpha=\alpha_s^{n,i}$ (red circles in
Figure~\ref {Fig:Wn0}).

The spectrum of a quantum mechanical problem determined by $V(\alpha ,y)$ is
given by the intersection of a vertical line through the chosen value of $%
\alpha $ and the blue lines in~Figure~\ref{Fig:Wn0} (obviously, of the
infinite number of the latter lines we only show $7$). Figure~\ref{Fig:Wn0}
shows two such vertical lines (green, dashed). It is worth noticing that any
such vertical line will meet only one red point when $\alpha =\alpha
_{s}^{(n,i)}$ and none when $\alpha \neq \alpha _{s}^{(n,i)}$ which tells us
that the truncation condition gives only one eigenvalue and for a particular
model potential. An exception should be made for the trivial case $\alpha =0$
because the truncation condition yields all the eigenvalues of the harmonic
oscillator. The reason is that the Frobenius method for the harmonic
oscillator leads to a two-term recurrence relation and it can be proved that
there are no square-integrable solutions beyond those with polynomial factors%
\cite{P68}. The origin of the authors' mistakes appears to be that they
think that the approach that gives the whole spectrum of the exactly
solvable model ($\alpha =0$) also gives the whole spectrum of the
conditionally-solvable one. It is already well known that such an assumption
is false\cite{CDW00,AF20,F20b,F20c,T16}.

The Rayleigh-Ritz variational method is extremely reliable and is
commonly used for obtaining the most accurate eigenvalues of
atomic and molecular systems\cite{P68,D20}. However, in order to
verify the accuracy of present results we have also applied the
powerful Riccati-Pad\'e method (RPM)\cite{FMT89a} that exhibits
exponential convergence. This approach is based on a rational
approximation to the power-series expansion of the logarithmic
derivative of the wavefunction ($f(y)$ in the present case) and
does not exhibit any feature common to the Rayleigh-Ritz
variational method\cite{FMT89a}. For this reason the RPM is a most
reliable and independent test for the results obtained by means of
the Rayleigh-Ritz method. A curious feature of the RPM is that,
given a value of $\alpha$, it yields the eigenvalues for $\pm
\alpha$. Tables \ref{tab:RPM1} and \ref{tab:RPM2} show that the
roots of the Hankel determinants\cite{FMT89a} converge towards the
eigenvalues given by the Rayleigh-Ritz variational method
discussed above (tables \ref{tab:alpha1} and \ref{tab:alpha2}) as
the determinant dimension $D$ increases. Notice that the RPM also
yields the exact eigenvalue $W=W_0^{(1)}=4$ for $\alpha=\pm
\sqrt{2}$.

Finally, we mention that the truncation condition yields only positive
eigenvalues $W_{s}^{(n)}$; however, the actual eigenvalues $W_{j,s}(\alpha )$
decrease with $\alpha $ and, eventually, become negative. For example, it is
not difficult to show that
\begin{equation}
W_{j,s}(\alpha )\approx -\frac{\alpha ^{2}}{\left( 2j+2s+1\right) ^{2}}+%
\mathcal{O}\left( \alpha ^{-2}\right) ,\;\alpha \rightarrow \infty .
\end{equation}
Notice that the variational results in Figure~\ref{Fig:Wn0} illustrate this
behaviour.

\textit{Summarizing}: the analytical expression for the energy
obtained by the Vieira and Bakke is unsuitable for any physical
purpose because any change of the quantum number $l$ transforms
the chosen model into another one with a different interaction
potential. The dependence of the oscillator frequency on $n$ and
$l$ is an artifact of the truncation of the Frobenius series and
exhibits no physical meaning whatsoever. Present variational
results already show that the eigenvalues are continuous functions
of the model parameters and, consequently, the angular frequency
can have any positive value. The truncation condition only
provides some rare eigenvalues and eigenfunctions with polynomial
factors $H_{s}^{(n,i)}(y)$ that by themselves do not represent the
spectrum of a single problem but particular solutions of more than
one model. Apparently, the authors were unaware of this fact. The
origin of the authors' misunderstanding of the solutions given by
the truncation method is the false belief that the eigenfunctions
with polynomial factors $H_{s}^{(n,i)}(y)$ are the only possible
solutions. Simple inspection of the eigenvalue equation
(\ref{eq:eigen_eq}), supported by a variational calculation,
already shows that most of the eigenfunctions do not exhibit
polynomial factors. The truncation condition only yields the whole
spectrum for the trivial case $\alpha =0$ and the reason is that
in this case the Frobenius method leads to a two-term recurrence
relation\cite {P68}. In this Comment we have also shown how to
extract some useful information from the roots of the truncation
condition.

\begin{table}[tbp]
\caption{Lowest variational eigenvalues
$W_{j,0}\left(-\sqrt{2}\right)$} \label{tab:alpha1}
\begin{center}
\par
\begin{tabular}{D{.}{.}{3}D{.}{.}{11}D{.}{.}{11}D{.}{.}{11}D{.}{.}{11}}
\hline \multicolumn{1}{c}{$N$}& \multicolumn{1}{c}{$W_{0,0}$} &
\multicolumn{1}{c}{$W_{1,0}$} & \multicolumn{1}{c}{$W_{2,0}$}  & \multicolumn{1}{c}{$W_{3,0}$}\\
\hline

 2  &  4.000000000  &   10.49997602  &                &                \\
 3  &  4.000000000  &   7.751061995  &   19.88102859  &                \\
 4  &  4.000000000  &   7.694010921  &   11.97562584  &   33.92039998  \\
 5  &  4.000000000  &   7.693979367  &   11.51212379  &   17.05520450  \\
 6  &  4.000000000  &   7.693978905  &   11.50604696  &   15.46896992  \\
 7  &  4.000000000  &   7.693978892  &   11.50604243  &   15.37652840  \\
 8  &  4.000000000  &   7.693978891  &   11.50604238  &   15.37592761  \\
 9  &  4.000000000  &   7.693978891  &   11.50604238  &   15.37592718  \\
10  &  4.000000000  &   7.693978891  &   11.50604238  &   15.37592718  \\
\end{tabular}
\par
\end{center}
\end{table}

\begin{table}[tbp]
\caption{Lowest variational eigenvalues
$W_{j,0}\left(\sqrt{2}\right)$} \label{tab:alpha2}
\begin{center}
\par
\begin{tabular}{D{.}{.}{3}D{.}{.}{11}D{.}{.}{11}D{.}{.}{11}D{.}{.}{11}}
\hline \multicolumn{1}{c}{$N$}& \multicolumn{1}{c}{$W_{0,0}$} &
\multicolumn{1}{c}{$W_{1,0}$} & \multicolumn{1}{c}{$W_{2,0}$}  & \multicolumn{1}{c}{$W_{3,0}$}\\
\hline

 2&  -1.180391283 &  4.000000000 &              &               \\
 3&  -1.401182256 &  4.000000000 &  9.284143096 &               \\
 4&  -1.449885589 &  4.000000000 &  8.345259771 &  17.66452696  \\
 5&  -1.458156835 &  4.000000000 &  8.344361267 &  12.69095166  \\
 6&  -1.459389344 &  4.000000000 &  8.344349784 &  12.53313315  \\
 7&  -1.459560848 &  4.000000000 &  8.344349442 &  12.53290257  \\
 8&  -1.459583736 &  4.000000000 &  8.344349427 &  12.53290132  \\
 9&  -1.459586704 &  4.000000000 &  8.344349427 &  12.53290130  \\
10&  -1.459587081 &  4.000000000 &  8.344349427 &  12.53290130  \\
11 &  -1.459587128 & 4.000000000 & 8.344349427 & 12.53290130  \\
12 &  -1.459587134 & 4.000000000 & 8.344349427 & 12.53290130  \\
13 &  -1.459587134 & 4.000000000 & 8.344349427 & 12.53290130  \\
\end{tabular}
\par
\end{center}
\end{table}

\begin{table}

\caption{Lowest RPM eigenvalues $W_{j,0}\left(-\sqrt{2}\right)$}

\label{tab:RPM1}

\begin{center}

\begin{tabular}{D{.}{.}{3}D{.}{.}{11}D{.}{.}{11}D{.}{.}{11}D{.}{.}{11}}
\hline \multicolumn{1}{c}{$D$}& \multicolumn{1}{c}{$W_{0,0}$} &
\multicolumn{1}{c}{$W_{1,0}$} & \multicolumn{1}{c}{$W_{2,0}$}  & \multicolumn{1}{c}{$W_{3,0}$}\\
\hline

 8  &  4.000000000  &   7.693449704  &   11.16567414  &                \\
 9  &  4.000000000  &   7.693990279  &   11.54430665  &   15.16153572  \\
10  &  4.000000000  &   7.693978617  &   11.50547430  &   15.12357546  \\
11  &  4.000000000  &   7.693978898  &   11.50605268  &   15.41617482  \\
12  &  4.000000000  &   7.693978891  &   11.50604217  &   15.37544063   \\
13  &  4.000000000  &   7.693978891  &   11.50604238  &   15.37593481  \\
14  &  4.000000000  &   7.693978891  &   11.50604238  &   15.37592704  \\
15  &  4.000000000  &   7.693978891  &   11.50604238  &   15.37592718  \\

\end{tabular}

\end{center}

\end{table}

\begin{table}

\caption{Lowest RPM eigenvalues $W_{j,0}\left(\sqrt{2}\right)$}

\label{tab:RPM2}

\begin{center}

\begin{tabular}{D{.}{.}{3}D{.}{.}{11}D{.}{.}{11}D{.}{.}{11}D{.}{.}{11}}
\hline \multicolumn{1}{c}{$D$}& \multicolumn{1}{c}{$W_{0,0}$} &
\multicolumn{1}{c}{$W_{1,0}$} & \multicolumn{1}{c}{$W_{2,0}$}  & \multicolumn{1}{c}{$W_{3,0}$}\\
\hline

 8  &  -1.459586733  &   4.000000000  &  8.413675510   &   12.38593112             \\
 9  &  -1.459587149  &   4.000000000  &  8.343322691   &   12.35721732             \\
10  &  -1.459587134  &   4.000000000  &  8.344372803   &   12.62125942  \\
11  &  -1.459587135  &   4.000000000  &  8.344348852   &   12.53183748  \\
12  &  -1.459587134  &   4.000000000  &  8.344349441   &   12.53292129    \\
13  &  -1.459587134  &   4.000000000  &  8.344349426   &   12.53290088    \\
14  &  -1.459587134  &   4.000000000  &  8.344349427   &   12.53290131   \\
15  &  -1.459587134  &   4.000000000  &  8.344349427   &   12.53290130  \\

\end{tabular}

\end{center}

\end{table}

\begin{thebibliography}{9}
\bibitem{VB20}  S. L. R. Vieira and K. Bakke, Phys. Rev. A \textbf{101},
032102 (2020).

\bibitem{F20}  F. M. Fern\'{a}ndez, Dimensionless equations in
non-relativistic quantum mechanics, arXiv:2005.05377 [quant-ph].

\bibitem{G32}P. G\"uttinger, Z. Phys. {\bf 73}, 169 (1932).

\bibitem{F39}  R. P. Feynman, Phys. Rev. \textbf{56}, 340 (1939).

\bibitem{CDW00}  M. S. Child, S-H. Dong, and X-G. Wang, J. Phys. A \textbf{33%
}, 5653 (2000).

\bibitem{AF20}  P. Amore and F. M. Fern\'{a}ndez, On some conditionally
solvable quantum-mechanical problems,arXiv:2007.03448 [quant-ph].

\bibitem{F20b}  F. M. Fern\'{a}ndez, The rotating harmonic oscillator
revisited, arXiv:2007.11695 [quant-ph].

\bibitem{F20c}  F. M. Fern\'{a}ndez, The truncated Coulomb potential
revisited, .arXiv:2008.01773 [quant-ph].

\bibitem{T16}  A. V. Turbiner, Phys. Rep. \textbf{642}, 1 (2016).
arXiv:1603.02992v2

\bibitem{P68}  F. L. Pilar, Elementary Quantum Chemistry (McGraw-Hill, New
York, 1968).

\bibitem{D20}G. W. F. Drake, J. Phys. B {\bf 53}, 223001 (2020).

\bibitem{FMT89a} F. M. Fern\'andez, Q. Ma, and R. H. Tipping, Phys. Rev. A {\bf 39}, 1605
(1989).
\end{thebibliography}
\end{document}